\documentclass[aps,prl,reprint,superscriptaddress]{revtex4-1}
\usepackage{graphicx}
\usepackage{amsmath}
\usepackage{amssymb}

\begin{document}
\title{Reply to ``Comment on `Vortex distribution in a confining potential ' "}

\author{Matheus Girotto}
\email{girotto.matheus@gmail.com}
\affiliation{Instituto de F\'isica, Universidade Federal do Rio Grande do Sul, Caixa Postal 15051, CEP 91501-970, Porto Alegre, RS, Brazil}

\author{Alexandre P. dos Santos}
\email{alexandreps@ufcspa.edu.br}
\affiliation{Departamento de Educa\c c\~ao e Informa\c c\~ao em Sa\'ude, Universidade Federal de Ci\^encias da Sa\'ude de Porto Alegre, 90050-170, Porto Alegre, RS, Brazil.}
\affiliation{Departamento de F\'isica, Universidade Federal de Santa Catarina, 88040-900, Florian\'opolis, Santa Catarina, Brazil}

\author{Renato Pakter}
\email{pakter@if.ufrgs.br}
\affiliation{Instituto de F\'isica, Universidade Federal do Rio Grande do Sul, Caixa Postal 15051, CEP 91501-970, Porto Alegre, RS, Brazil}

\author{Yan Levin}
\email{levin@if.ufrgs.br}
\affiliation{Instituto de F\'isica, Universidade Federal do Rio Grande do Sul, Caixa Postal 15051, CEP 91501-970, Porto Alegre, RS, Brazil}

\begin{abstract}
We argue that contrary to recent suggestions, non-extensive statistical mechanics has no relevance
for inhomogeneous systems of particles interacting by short-range potentials. We show that these
systems are perfectly well described by the usual Boltzmann-Gibbs statistical mechanics.
\end{abstract}

\maketitle

In a recent Phys. Rev. Lett.~\cite{AnDa10}, Andrade \textit{et al.} studied a system of particles (vortices) 
interacting by the potential 
\begin{equation}\label{V}
V(r)=q^2 K_0(\dfrac{|{\bf x}_1-{\bf x}_2|}{\lambda})
\end{equation}
where $K_0$ is a modified Bessel function, $r=|{\bf x}_1-{\bf x}_2|$ is the distance between particle~$1$ and particle~$2$, $q$ is the potential strength, and $\lambda$ is the screening length.  The particles were confined to a potential well 
\begin{equation}\label{w}
W(x)=\alpha \frac{x^2}{2} \ .
\end{equation}
The principal conclusion of the paper by Andrade \textit{et al.} was that a system of such particles, in contact with a reservoir at $T=0$, ``obeys Tsallis statistics".  At finite temperatures, the authors argued,
that the system will maximize a mixture of Tsallis and Bolzmann entropy.  In our Comment~\cite{LePa11} on Andrade \textit{et al.} paper, we pointed out that at $T=0$, statistics is irrelevant and a system in contact with a reservoir 
at $T=0$ will loose all of its free energy and will collapse into the ground state.  We then explicitly 
calculated the particle distribution in the ground state in the limit $N \rightarrow \infty$, $q \rightarrow 0$, and
$N q^2=1$, and showed that it is different from the one predicted 
by Tsallis entropy.  

In the follow up paper~\cite{GiDo13}, we have extended our theory to finite temperatures and showed how the
system of Andrade \textit{et al.} can be studied using a mean-field theory. The Comment of Ribeiro \textit{et al.}~\cite{RiNo14} criticizes our paper and insists that the equilibrium state of the 
system, described by Eqs.~\eqref{V} and \eqref{w}, should be
described by the non-extensive statistical mechanics.

Below we address the issues raised by Ribeiro \textit{et al.}:

\begin{enumerate} 

\item The asymptotic form of the interaction potential in Eq.~\eqref{V} is $V(r) \approx q^2 \sqrt{\frac{\pi}{2 r}} e^{-r/\lambda}$.  This potential is short-ranged and has a form very similar
to Yukawa potential.  It is well known that a system of Yukawa particles confined by hard walls or
periodic boundary conditions crystallizes~\cite{RoKr88,MeFr91,LoPa93,StRo93,HoRo04,GaNa12}. The process is perfectly well described by the standard 
Boltzmann-Gibbs~(BG) statistical mechanics.   Ribeiro \textit{et al.} do not
provide any argument why the equilibrium state of the Yukawa-like system confined by
a parabolic potential should be 
described by a non-extensive entropy.  The only arguments are based on fitting the particle
distributions calculated using overdamped Molecular Dynamics~(MD) simulations to q-Gaussians.
Such curve fitting, however, must be taken with a grain of salt.  For example, recently,
it has been argued that sufficiently strongly correlated random variables also obey ``non-extensive" 
central limit theorem in which the usual Gaussian distribution for uncorrelated random variables 
is replaced by a q-Gaussian.  Again the only basis for 
this belief was curve fitting.  However, in an important paper, Hilhorst and Schehr~\cite{HiSc07} calculated  exactly the probability distributions for strongly correlated random variables and showed that these are analytically different from the q-Gaussians. Curve fitting is a very shaky ground on which to build a new theory, in particular the one that attempts to replace the BG statistical mechanics.

\item In their Comment on our work the authors state that 
``besides the long-range forces, other attributes, like strong correlations" make systems 
fall out of the ``scope
of BG statistical mechanics".  Indeed some years ago, it was hoped that the non-extensive statistics could be helpful to study systems 
with {\it long-range} interactions, such as magnetically confined plasmas or gravitational clusters. 
However, recent work~\cite{CaDa09,LePa14} has shown that long-range interacting systems relax to quasi-stationary states which have nothing to do with Tsallis entropy.

It is also incorrect to say that BG statistics
fails for strongly correlated systems.  If this would be true, the theory could not be used to study either liquids or solids, which are very strongly correlated.  Yet, BG statistical mechanics is able to account perfectly for
the structural and thermodynamic properties of liquid and solid phases, as well as for the phase transitions between the different phases.

\item  In their Comment on our paper, Ribeiro \textit{et al.} claim that we 
did not ``realize how poor mean-field approximation" was in the strong coupling regime.
The discrepancy between the mean-field and  MD simulations
at low temperatures was clearly pointed out by us.  Furthermore, it is very well know that 
the mean-field theory fails when the correlations between the 
particles become strong, see, for example, discussion in Ref.~\cite{Le02}.
The failure of the mean-field theory, however, is in no way indicative of the failure of the BG statistical 
mechanics.  Indeed, it is possible to solve ``exactly" for the particle distribution predicted by the BG
statistical mechanics using Monte Carlo~(MC).  This is precisely what we have done in our paper~\cite{GiDo13} (Fig.~5a from our paper).  We compared the results of overdamped dynamics simulations of Andrade \textit{et al.} with the the predictions of BG statistical mechanics for the same number of particles, the same parameters, and the same temperatures as in their paper.
As expected the results of Andrade \textit{et al.} are in perfect agreement with the predictions of BG statistical mechanics.  Ribeiro et al., say that the reason for this
good agreement is that the temperatures that
we looked at was ``too high".  They argue that at lower temperatures, more relevant for supperconductors, BG statistics will fail and the density distribution will correspond to the
maximum of Tsallis entropy.  To answer this criticism, in this Reply we calculated 
a coarse grained density distribution at the lowest temperature, $T=0$. 
Within usual thermodynamics and  BG statistical mechanics 
a system at $T=0$ will be in the ground state, which can be calculated by 
minimizing the potential energy of the system.  The coarse grained density distribution
is then constructed by binning the particles along the x-axis.  In Fig. 1 we show, that
this density distribution is in excellent agreement with the overdamped dynamics data of Andrade et al. 
Once again we conclude that there is absolutely no reason to introduce
a non-extensive entropy for the system of particles interacting by a short-range potential.

\begin{figure}
\begin{center}
\includegraphics[width=0.5\textwidth]{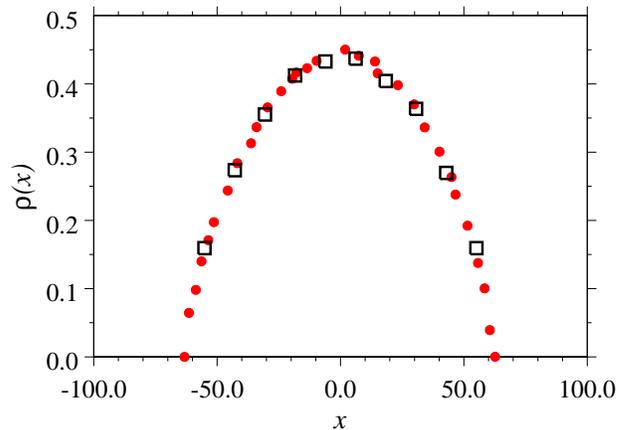}
\caption{ Comparison between the overdamped dynamics data of Andrade \textit{et al.} \cite{AnDa10} (cicles) 
with the predictions of BG statistics (squares) for $T=0$.  The perfect agreement between the two clearly shows that the equilibrium state of the system of Andrade \textit{et al.} is described by the standard BG statistical mechanics down to $T=0$. \label{fig}}
\end{center}
\end{figure}

\item Ribeiro \textit{et al.} claim that their W-Lambert solution describes perfectly the MD data of Andrade \textit{et al.}
However, they fail to point out that the good agreement shown in the Fig. 1 of their Comment 
is obtained with the help of a fitting 
parameter $a$.   Clearly if one has to decide between two very distinct theories 
which equally well account for the data, but one of which has a fitting parameter and the other one does not,
there is no question which theory should to be preferred.

\end{enumerate}

We showed 
that all of the overdamped dynamics data of Andrade et al, including $T=0$, is
perfectly well described by the BG statistical mechanics.  At high temperatures, the 
particle distribution can be accurately calculated using the mean-field theory.  At intermediate 
temperatures the correlations can be included  using a density functional theory in conjunction with the Hypernetted Chain Equation~(HNC)~\cite{GiDo14}.  Therefore, there is absolutely no reason to introduce a
non-extensive entropy for this problem.

This work was partially supported by the CNPq, FAPERGS, 
INCT-FCx, and by the US-AFOSR under the grant 
FA9550-12-1-0438.

\bibliography{ref.bib}

\end{document}